\documentclass[aps,pre,showpacs,twocolumn]{revtex4}
\usepackage{graphicx}
\usepackage[dvips]{epsfig}
\usepackage{dcolumn}
\usepackage{bm}
\usepackage{amssymb}
\usepackage{amsmath}

\begin{document}
\title{Comment on ``Mean First Passage Time for Anomalous Diffusion''}
\author{S.  B. Yuste}
\affiliation{Departamento de F\'{\i}sica, Universidad de Extremadura,
E-06071, Badajoz, Spain}

\author{Katja Lindenberg}
\affiliation{Department of Chemistry and Biochemistry, and Institute for
Nonlinear Science, University of California San Diego,  La Jolla, CA
92093-0340, USA}

\begin{abstract}
We correct a previously erroneous calculation
[Phys. Rev. E {\bf 62}, 6065 (2000)] of the mean first passage
time of a subdiffusive process to reach either end of a finite interval in
one dimension.  The mean first passage time is in fact infinite.
\end{abstract}

 \maketitle

Anomalous diffusion is
commonly characterized by the behavior of the mean squared displacement
as a function of time at long times\cite{Kehr,HilferEd,MetKlaPhysRep},
\begin{equation}
\langle x^2 \rangle \sim \frac{2 K_\alpha}{\Gamma(1+\alpha)} t^\alpha
\label{msd}
\end{equation}
where $K_\alpha$ is a generalized diffusion constant.
Ordinary diffusion corresponds to $\alpha=1$ and, in the more usual
notation, $K_1=D$.  {\em Superdiffusion} is associated with motion that
is faster than diffusive, $\alpha>1$, while {\em subdiffusion} occurs when
$\alpha<1$. 
In a recent paper~\cite{Gitterman}, the mean first passage times to the
ends of an interval for a
superdiffusive and a subdiffusive random walker on a line were
calculated, and these results have even more recently been applied to
the problem of anomalous heat conduction in such a line in the presence
of a temperature gradient~\cite{Li}.  However, there is an
error in the calculation for the {\em subdiffusive} problem from which
one concludes that the mean first passage time reported
in~\cite{Gitterman} for this case are incorrect.  {\em In fact, the mean first
passage time for the subdiffusive problem is infinite.}

To support this observation it is compelling to note that a
continuous time subdiffusive nearest neighbor
random walk (CTRW)~\cite{MetKlaPhysRep} with a waiting time
distribution which has a long tail, i.e., a walk in which the
probability density that a particle takes the next step at a time $t\to\infty$
after the previous step is
$\psi(t)\sim C_\alpha/t^{1+\alpha}$
leads exactly to Eq.~(\ref{msd}) ($C_\alpha$ is a constant). The mean
time for the particle to make
{\em a single} jump is
\begin{equation}
T_1= \lim_{t\to\infty} T_1(t),
\end{equation}
where
\begin{equation}
T_1(t)= \int_0^t \tau\psi(\tau){\mathrm d}\tau \label{t1a}.
\end{equation}
For large $t$,
\begin{equation}
T_1(t) \propto \int^t \frac{{\mathrm d}\tau}{\tau^\alpha} \propto
t^{1-\alpha}, \label{t1}
\end{equation}
so that one obtains the well-known result $T_1=\infty$,
i.e., the mean time to go from any one location to another
even in a single jump is infinite.  However,
this argument might generate issues about the waiting time
for the first step of the process, since one of the differences between
a CTRW and the fractional diffusion equation lies precisely in the
assumptions associated with this first step.  In a CTRW there is a
singular contribution to the probability density that the particle is
still at the origin $x_0$ at time $t$,
\begin{equation}
P(x,t)\sim \frac{C_\alpha}{\alpha}t^{-\alpha} \delta(x_0)
+~~\mbox {other terms},
\end{equation}
which does not appear in the solution of the fractional diffusion
equation~\cite{Barkai}.

To sidestep this problem and show that the divergence of the mean first
passage time does not arise only from this term, we also obtain the
divergent result starting with the fractional diffusion
equation that was the starting point in~\cite{Gitterman}. Although
his general formulation was for an arbitrary starting site in the
interval $(0,L)$ and in the presence of an external constant
force, the explicit final result was presented for a particular
initial location, $x=L/2$, and with no external force.  This
explicit result is also the one used in~\cite{Li}. We thus
restrict our presentation to this specific case.

The mean first passage time from $x=L/2$ to either $x=0$ or $x=L$
is given by~\cite{Gardiner}
\begin{equation}
T=\int_0^L {\mathrm d}x \int_0^\infty {\mathrm d}t P(x,t)=
\int_0^\infty {\mathrm d}t S(t), \label{mfptg}
\end{equation}
where
\begin{equation}\label{St}
S(t)=\int_0^L {\mathrm d}x P(x,t)
\end{equation}
and  $P(x,t)$ is the solution of the fractional diffusion
equation~\cite{MetKlaPhysRep,Scheider,West}
\begin{equation}
\frac{\partial}{\partial t} P(x,t) = K_\alpha \; _0D^{1-\alpha}_t
\frac{\partial^2}{\partial x^2} P(x,t)
\label{fde}
\end{equation}
with absorbing boundary conditions
$P(0,t)=P(L,t)=0$
and initial condition
$P(x,t=0)=\delta(x-L/2)$.
Here $_0D^{1-\alpha}_t$ is the Riemann-Liouville operator
\begin{equation}
_0D^{1-\alpha}_t P(x,t) = \frac{1}{\Gamma(\alpha)}
\frac{\partial}{\partial t} \int_0^t {\mathrm d}\tau
\frac{P(x,\tau)}{(t-\tau)^{1-\alpha}}
\end{equation}
and $K_\alpha$ is the generalized diffusion coefficient in
Eq.~(\ref{msd}).  The quantity $S(t)$ is called the survival
probability  because, as one sees from Eq.~\eqref{St}, $S(t)$ is
just the probability that the particle has not been absorbed by
the boundaries at $x=0$ and $x=L$ during the time interval
$[0,t]$.

The solution of Eq.~(\ref{fde}) with the given boundary and initial
conditions can be found by the method of separation of
variables~\cite{MetKlaPhysRep,rangding}:
\begin{eqnarray}
P(x,t)&=&\frac{2}{L} \sum_{n=0}^\infty (-1)^n \sin\frac{(2n+1)\pi
x}{L} \nonumber\\
&& \times E_\alpha\left(-K_\alpha(2n+1)^2\pi^2t^\alpha/L^2\right).
\end{eqnarray}
Here $E_\alpha(-z)$ is the Mittag-Leffler function (for $\alpha=1$
it reduces to the exponential $\exp(-z)$ and thus yields the usual
solution for the diffusive problem). It then follows that
\begin{equation}
S(t)=\frac{4}{\pi} \sum_{n=0}^\infty \frac{(-1)^n}{2n+1}
E_\alpha(-K_\alpha (2n+1)^2\pi^2t^\alpha/L^2). \label{st}
\end{equation}
The mean first passage time to $x=0$ or $L$ is then $T=
\lim_{t\to\infty} T(t)$ where
\begin{equation}
T(t)= \int_0^t {\mathrm d}\tau S(\tau). \label{mfpt}
\end{equation}

To address the convergence of $T(t)$ for $t\to\infty$ we need to
analyze the long-time behavior of $S(t)$.  Note that $S(t)$ is
well behaved for finite times (the survival probability
goes to 1 for $t\rightarrow 0$), so that the divergence of $T$
is due to the behavior
at long times. For large $z$ the Mittag-Leffler function behaves
as
\begin{equation}
E_\alpha(-z) \sim \sum_{m=1}^\infty
\frac{(-1)^{m+1}}{\Gamma(1-\alpha m)} z^{-m},
\end{equation}
and consequently, for large $t$,
\begin{eqnarray}
S(t)&\sim& \frac{4}{\pi} \sum_{n=0}^\infty \frac{(-1)^n}{2n+1}
\sum_{m=1}^\infty\frac{(-1)^{m+1} L^{2m}}{\Gamma(1-\alpha m)
[K_\alpha
(2n+1)^2\pi^2t^\alpha]^m}\nonumber\\
&\sim& \sum_{m=1}^\infty \frac{(-1)^{m+1} L^{2m}} {\Gamma(1-\alpha
m) [\pi^2K_\alpha t^\alpha]^m} Z(m),
\end{eqnarray}
where $Z(m)=\sum_{n=0}^\infty (-1)^n/(2n+1)^{2m+1}$. For
$t\to\infty$, and using the fact that $Z(1)=\pi^3/32$, we
then have
\begin{equation}
S(t) \sim \frac{1}{8\Gamma(1-\alpha)} \frac{L^2}{K_\alpha
t^\alpha} .
\end{equation}
It then follows that for large $t$ we have
\begin{equation}
T(t) =\int_0^t {\mathrm d}\tau S(\tau)  \sim
\frac{1}{8(1-\alpha)\Gamma(1-\alpha)} \frac{L^2}{K_\alpha
t^{\alpha-1}},
\end{equation}
i.e., $T(t)\propto t^{1-\alpha}$, exactly as in Eq.~(\ref{t1}). We
thus conclude that $T(t) \to \infty$ when $t\to\infty$ for any
$\alpha<1$.

We have thus shown that the mean first passage time
for a subdiffusive process described by the fractional diffusion equation
(or, for that matter, by a continuous time random walk) to reach the boundaries
of a one-dimensional interval is infinite. We note that our Eq.~\eqref{st}
appears as Eq.~(40) in ~\cite{metzklaf}, but the connection between the
survival probability and the mean first passage time is never made in that
work so a user of the result in~\cite{Gitterman} would not necessarily
discover the connection.  An expression for the first passage time density
involving the Mittag-Leffler function appears as Eq.~(3.87)
in~\cite{rangding}, but, again, they do not go on to calculate the mean
first passage time, nor do they do the necessary asymptotic analysis of the
Mittag-Leffler function that would allow them to do so.  That these results
would not necessarily lead a reader to conclude that the mean first passage
time is infinite is reinforced by the fact that both of these references
appear in~\cite{Gitterman}.

Finally, we note that the validity of the result in~\cite{Gitterman}
for the mean first passage time in the superdiffusive regime has
also  been questioned recently because it violates a theorem due to
Sparre Andersen~\cite{chechkin,andersen}.

\section*{Acknowledgments}

This work has been partially supported by the Ministerio de
Ciencia y Tecnolog\'{\i}a (Spain) through Grant No. BFM2001-0718
and by the Engineering Research Program of the Office of Basic
Energy Sciences at the U. S. Department of Energy under Grant No.
DE-FG-86ER13606.

\end{document}